\documentclass[amsmath,nofootinbib,amssymb]{revtex4}
\pdfoutput=1
\usepackage[utf8]{inputenc}
\usepackage{float}
\usepackage{epsfig}
\usepackage{graphicx}
\usepackage{dcolumn}
\usepackage{bm}
\usepackage{color}
\usepackage{tabularx}
\usepackage{cleveref}
\newcommand{\av}[1]{\left\langle #1 \right\rangle}

\newcommand{\di}{\text{d}}
\newcommand{\n}{\phantom{a}}
\usepackage{subfigure}

\newcommand{\lsim}{\mbox{\raisebox{-.9ex}{~$\stackrel{\mbox{$<$}}{\sim}$~}}}
\newcommand{\gsim}{\mbox{\raisebox{-.9ex}{~$\stackrel{\mbox{$>$}}{\sim}$~}}}

\def\thebiblio#1{
\begin{center}\bf \large References
\end{center}
\list
{[\arabic{enumi}]}{\settowidth\labelwidth{#1.}\leftmargin\labelwidth
 \advance\leftmargin\labelsep
 \usecounter{enumi}}
 \def\newblock{\hskip .11em plus .33em minus -.07em}
 \sloppy
 \sfcode`\.=1000\relax}
 
\usepackage{color}
\usepackage{verbatim}

\begin{document}
\preprint{}
\title{Constraints on anharmonic corrections of Fuzzy Dark Matter}

\author{J.\,A.\,R.\,Cembranos\footnote{cembra@ucm.es}, 
        A.\,L.\,Maroto\footnote{maroto@ucm.es},
        S.\,J.\,N\'u\~nez Jare\~no,
        and H.\,Villarrubia-Rojo\footnote{hectorvi@ucm.es}}
\address{Departamento de  F\'{\i}sica Te\'orica, Universidad Complutense de Madrid, E-28040 Madrid, Spain}
\date{\today}

\begin{abstract}
    The cold dark matter (CDM) scenario has proved successful in cosmology. However, we lack a 
    fundamental understanding of its microscopic nature. Moreover, the apparent disagreement between
    CDM predictions and subgalactic-structure observations has prompted the debate about its behaviour
    at small scales.
    These problems could be alleviated if the dark matter is composed of ultralight fields 
    $m\sim 10^{-22}\ \text{eV}$, usually known as fuzzy dark matter (FDM). Some specific models, 
    with axion-like potentials, have been thoroughly studied and are collectively referred to as 
    ultralight axions (ULAs) or axion-like particles (ALPs). In this work we consider anharmonic 
    corrections to the mass term coming from a repulsive quartic self-interaction. Whenever this 
    anharmonic term dominates, the field behaves as radiation instead of cold matter, modifying the time 
    of matter-radiation equality. Additionally, even for high masses, i.e. masses that reproduce the cold 
    matter behaviour, the presence of anharmonic terms introduce a cut-off in the matter power spectrum 
    through its contribution to the sound speed. We analyze the model and derive constraints using a 
    modified version of {\sc class} and comparing with CMB and large-scale structure data.
\end{abstract}

\maketitle

\section{Introduction}\label{sec:intro}
    The evidence collected over the last decades suggests that most of the matter in the universe 
    exists in the form of dark matter (DM), whose effects have only been detected through its 
    gravitational interaction. In particular, the assumption that dark matter is composed of 
    non-relativistic particles, the so-called cold dark matter (CDM), has produced a remarkable 
    concordance with the observational data over a wide range of scales and evolution epochs. It is one
    of the foundations of the succesful standard cosmological model $\Lambda$CDM.\\
    
    Notwithstanding agreement with observations, several ingredients are lacking in our understanding
    of DM. In the first place, we have been unable to detect any non-gravitational interaction of DM.
    Most of the work in the field is currently devoted to direct, indirect detection and production
    searches. Owing to this effort it has been possible to tighten the parameter space of the most
    popular models. This lack of additional interactions makes it more difficult to discriminate 
    between different models. There are many candidates that behave like CDM on cosmological scales, with 
    masses ranging from the meV of the QCD axion \cite{Graham:2015ouw} to the TeV \cite{Goldberg:1983nd, Ellis:1983ew} 
    and going up to the 100 $M_\odot$ of the primordial black holes \cite{Carr:2016drx}. 
    The other ingredient missing is a precise understanding of the DM behaviour on small, i.e. galactic,
    scales. Even though most DM models mimic CDM on cosmological scales, their predictions usually 
    differ on smaller scales \cite{Ostriker:2003qj} so they could be discriminated based only on their 
    gravitational effects. In fact, there exist three long-standing debates, questioning the agreement
    between observations and the CDM theoretical predictions \cite{Weinberg:2013aya, Pontzen:2014lma}, 
    the so-called `\emph{too big to fail}' \cite{BoylanKolchin:2011de}, `\emph{missing satellites}'
    \cite{Moore:1999nt} and especially the `\emph{core-cusp}' problem \cite{deBlok:2009sp}. The 
    `\emph{core-cusp}' problem refers to the discrepancy between the density profiles of CDM
    halos obtained in $N$-body simulations, that tend to be cuspy in the center, and the ones inferred 
    from observations, that point to the existence of a central core. Although these problems are
    sometimes attributed to baryonic effects unaccounted for in the simulations \cite{Onorbe:2015ija}, 
    they remain one of the main challenges of the CDM model.\\
    
    An interesting alternative that neatly solves the `\emph{core-cusp}' problem is fuzzy dark
    matter (FDM) \cite{Hu:2000ke}. In this picture, dark matter is composed of ultralight particles with
    $m\sim 10^{-22}\;\text{eV}$, so that its Compton wavelength ($m^{-1}$) reaches astrophysical scales.
    Then, the formation of cusps is prevented \cite{Schive:2014dra}. The wave nature of the particles on 
    the smallest scales makes them impossible to localize.
    While solving this problem, FDM behaves as a rapidly oscillating coherent scalar field, thus
    recovering the CDM behaviour on cosmological scales. In his groundbreaking work \cite{Turner:1983he}, 
    Turner analyzed a homogeneous oscillating scalar field in an expanding universe. He showed that a 
    rapidly oscillating scalar field with a power-law potential $V(\phi)\propto \phi^n$ behaves as
    a perfect fluid with an effective equation of state $w=(n-2)/(n+2)$. More general expressions
    can be obtained from a version of the virial theorem \cite{Johnson:2008se}. The results of Turner show that
    a massive scalar field, i.e. harmonic potential, oscillating coherently with a frequency much higher
    than the expansion rate behaves as CDM, at least at the background level. Afterwards, ultralight 
    scalar fields have been thoroughly studied at the perturbation level \cite{Johnson:2008se, 
    Hwang:2009js, Park:2012ru, Hlozek:2014lca, Cembranos:2015oya}, proving that the same
    conclusion holds. Perturbations of coherent oscillating scalar fields admit an effective fluid 
    description with an effective sound speed nearly zero, like CDM. The main cosmological signature
    of these models is the supression of growth at small scales. Below some Jeans scale $k_J^{-1}$ the modes
    do not grow appreciably, translating into a cut-off in the matter power spectrum \cite{Hlozek:2014lca}.
    Although the work on ultralight fields has been mainly concerned with scalar fields, there are
    recent results on higher spin fields. It has been shown that abelian vectors at the background
    \cite{Cembranos:2012kk} and perturbation level \cite{Cembranos:2016ugq}, 
    non-abelian vectors \cite{Cembranos:2012ng} and arbitrary-spin fields \cite{Cembranos:2013cba} 
    behave in a similar way. Interestingly, the results of \cite{Cembranos:2013cba}
    show that it is possible to achieve an isotropic model of higher-spin dark matter as long as it is
    rapidly oscillating.\\
    
    These ideas have been applied to the axion, a particularly well-motivated DM candidate. The standard
    QCD axion was initially proposed to solve the strong $CP$ problem \cite{Peccei:1977hh, Wilczek:1977pj,
    Weinberg:1977ma} in particle physics. Likewise, the appereance of many light scalar fields seems to 
    be a generic feature of different string-theory scenarios. Some of these fields have a similar origin 
    as the QCD axion, arising from the breaking of an approximate shift symmetry, and are usually known 
    as axion-like particles (ALPs) or ultralight axions (ULAs) \cite{Marsh:2015xka, Hui:2016ltb}. ALPs 
    present similar periodic potentials but with a mass much smaller than the QCD axion that could lie 
    in the range of ultralight fields $m\sim 10^{-22}\;\text{eV}$. While behaving like FDM, ALPs have 
    a rich phenomenology based on their assumed interaction with matter. Aside from the standard searches 
    for axions, there is a wealth of dedicated searches and projected experiments on the lookout for 
    ultralight axions. These include studies of the neutral hydrogen distribution in the universe 
    \cite{Sarkar:2015dib,  Kobayashi:2017jcf}, laboratory constraints based on nuclear interactions 
    \cite{Abel:2017rtm}, astrophysical bounds \cite{Banik:2017ygz, Hirano:2017bnu, Conlon:2017ofb}, 
    gravitational wave searches \cite{Brito:2017wnc, Brito:2017zvb} and analysis of CMB spectral 
    distortions \cite{Sarkar:2017vls, Diacoumis:2017hff}. A prominent feature of the model is the presence
     of anharmonic corrections over the mass term in FDM. These corrections arise, to first order, as 
    quartic corrections in the potential with the opposite sign of the mass term, i.e. attractive 
    self-interactions. These effects have been studied, as well as the effect of the full 
    axion potential \cite{Urena-Lopez:2015gur, Cedeno:2017sou} and their effect on the linear matter 
    power spectrum seems to be negligible. However, self-interactions could modify non-linear structures 
    in a significant way \cite{Desjacques:2017fmf}\\
    
    Another possibility involves introducing a positive quartic correction, i.e. repulsive 
    self-interactions. It is more difficult to find particle-physics models in this case 
    \cite{Fan:2016rda}, but the model is nonetheless well motivated as the simplest modification leading 
    to a stable potential. This modification has been previously analyzed in some works 
    \cite{Goodman:2000tg, Li:2013nal, Suarez:2016eez, Dev:2016hxv, Li:2016mmc}.
    The additional source of pressure from the repulsive self-interactions helps to solve 
    the `\emph{core-cusp}' problem with larger masses \cite{Fan:2016rda}. Additionaly, unlike the axion 
    case, it could explain the formation of vortices in galaxies \cite{RindlerDaller:2011kx}.\\
    
    In this work we will consider a fuzzy dark matter model with an additional quartic self-interaction.
    Using a modified version of the cosmological Boltzmann code {\sc class} \cite{Blas:2011rf} and 
    parameter-extraction code MontePython \cite{Audren:2012wb} we will constrain the parameters of the 
    model with CMB \cite{PlanckResults} and large-scale structure (LSS) \cite{WiggleZResults} data. 
    Section \ref{sec:exact} presents the model and
    the relevant equations for background and perturbation evolution. In section \ref{sec:avg}, we
    review the averaging procedure when the field is rapidly oscillating and the effective fluid
    equations in this case. Section \ref{sec:heur} discusses a simplified model and estimates analytic
    bounds on the parameters, highlighting the main physical effects and the origin of the constraints 
    on the model. In section \ref{sec:numeric} we present the result of the full numerical analysis and
    the final constraints on the model, as well as a discussion of its physical origin. Section \ref{sec:conclusions} summarizes the conclusions and 
    prospects for future work.
        
\section{Exact evolution}\label{sec:exact}
    Let us assume a scalar field with Lagrangian
	\begin{equation}
	    \mathcal{L} = \frac{1}{2}g^{\mu\nu}\partial_\mu  \phi\partial_\nu \phi - V(\phi),
	\end{equation}
	and potential
	\begin{equation}
	    V(\phi) = \frac{1}{2}m^2\phi^2 + \frac{1}{4}\lambda\phi^4\ ,
	\end{equation}
	in a homogeneous and isotropic universe with a flat Robertson-Walker metric in conformal time $\eta$
	\begin{equation}
        	\di s^2 = a^2(\eta) \left(\di\eta^2 - \di\bm{x}^2\right)\ .
	\end{equation}
	
	The equation of motion for a homogeneous scalar field in this background is
    \begin{equation}\label{eq:sf_bg}
		\ddot{\phi}+2\mathcal{H}\dot{\phi}+a^2 V'(\phi) = 0\ ,
	\end{equation}
	where $\mathcal{H} = \dot{a}/a$ and $\;\;\dot{}\equiv \partial/\partial \eta\;$.
	We choose initial conditions
	\begin{align}
	    \phi &= \phi_0\ ,\\
	    \dot{\phi} &= 0\ ,
	\end{align}
	where $\phi_0$ is chosen to match the desired energy density $\Omega_\phi$ today. These are the usual
	initial conditions when the axion-like particles are produced through a misalignment mechanism 
	\cite{Diez-Tejedor:2017ivd} and the field starts its evolution frozen. It is important to note that
	the choice of initial conditions has a deep impact in the subsequent evolution. In \cite{Li:2013nal},
	the authors considered a case similar to ours, but with an initial velocity $\dot{\phi}\neq 0$. In 
	this case, there is an initial phase of stiff-matter domination, absent in our case, constrained to 
	be short enough not to spoil BBN.
	
    We now introduce scalar perturbations over a 	flat Robertson-Walker metric. Following the notation of
    \cite{Mukhanov:1990me}, the general form of the perturbations is
    \begin{equation}
        \di s^2 = a^2(\eta)\Bigg[(1+2\Phi)\di\eta^2 - 2\partial_iB\di x^i\di\eta 
            - \Big((1-2\Psi)\delta_{ij}+2\partial_i\partial_jE\Big)\di x^i\di x^j\Bigg]\ .
    \end{equation}
	
	The equation of motion for the scalar field perturbation is
	\begin{equation}\label{eq:sf_pert}
	    \ddot{\delta\phi}+2\mathcal{H}\dot{\delta\phi}+(k^{2}+a^{2}V'')\delta\phi = 
	        \Big(k^2(B-\dot{E})+\dot{\Phi}+3\dot{\Psi}\Big)\dot{\phi}-2a^2\Phi V'\ ,
	\end{equation}
	where the gauge has not yet been fixed.
	We can introduce a different parameterization, reminiscent of a perfect fluid. The components of 
	the	perturbed energy-momentum tensor are \cite{Mukhanov:1990me}
	\begin{align}
		\delta T^0_{\n 0} &\equiv\delta\rho_\phi 
		    = a^{-2}(\dot{\phi}\dot{\delta\phi}-\dot{\phi}^{2}\Phi)+V'\delta\phi\ , \label{eq:delta_rho_phi}\\
		\delta T^i_{\n j} &\equiv -\delta P_\phi\,\delta^i_j 
		    = -\Big(a^{-2}(\dot{\phi}\dot{\delta\phi}-\dot{\phi}^{2}\Phi)-V'\delta\phi\Big)\,\delta^i_j\ , \label{eq:delta_P_phi}\\
		\delta T^0_{\n i} &\equiv (\rho_\phi +P_\phi)(v^i_\phi -\partial_i B)
		    =a^{-2}\dot{\phi}\, \partial_i\delta\phi\ . \label{eq:u_phi}
	\end{align}
	We can rewrite \eqref{eq:sf_pert} in terms of the fluid variables, introducing $\delta=\delta \rho/\rho$
	and $u=(1+w)(v-B)$. In the synchronous gauge, the metric variables read
	\begin{align*}
	    \Psi &= -\frac{1}{6}\big(h-\nabla^2\mu\big)\ ,\\
	    E &= \frac{1}{2}\mu\ ,\\
	    \Phi &= B = 0\ ,
	\end{align*}
	and the equations of motion are
	\begin{align}
		\dot{\delta} &= -3\mathcal{H}(1-w)\delta -ku-9\mathcal{H}^{2}(1-c^{2}_\text{ad})\frac{u}{k}
		    -\frac{1}{2}(1+w)\dot{h}\ , \label{eq:pert_d_exact}\\
		\dot{u} &= 2\mathcal{H}u +k\delta +3(w-c^{2}_\text{ad})\mathcal{H}u\ , \label{eq:pert_u_exact}
	\end{align}
	where $w=P/\rho$ is the equation of state and the adiabatic sound speed $c^2_\text{ad}$ is
	\begin{equation}
	    c^{2}_\text{ad}=\frac{\dot{P}}{\dot{\rho}}=1+\frac{2}{3}\frac{a^{2}V'}{\mathcal{H}\dot{\phi}}\ .
	\end{equation}
	
    Following the analysis of \cite{Hlozek:2014lca} we provide the system with initial conditions
	\begin{align}
		\delta &= 0\ ,\\
		u &= 0\ ,
	\end{align}							
	valid up to corrections of order $(k\eta)^4$.  The scalar field starts its evolution frozen in a value
	$\phi_0$ with an equation of state $w\simeq-1$. As the universe expands the field starts rolling down 
	the potential, when it reaches the minimum it undergoes rapid oscillations.
    These oscillations occur when the effective frequency $\omega_\text{eff}\sim\sqrt{V''(\phi)}$ is 
    bigger than the friction term $\mathcal{H}$, so once the scalar field starts oscillating its 
    frequency becomes much larger than the expansion rate, the inverse of the evolution time scale of 
    the background.
    
    On the numerical side, this means that it becomes prohibitely expensive to compute the exact evolution
    of the field, following every oscillation. However, the huge difference between time scales allows
    us to average the equations of motion and turn to an effective description.
	
\section{Averaged evolution}\label{sec:avg}    
    The study of the cosmological evolution of a fast oscillating scalar was first performed in 
    \cite{Turner:1983he}. Basically, if the oscillation frequency of the scalar field is much higher than 
    the expansion rate of the universe, the cosmological evolution becomes independent of the periodic 
    phase of the field at leading order. Consequently, the Einstein equations can be approximately solved
    averaging in time the energy-momentum tensor
	\begin{equation}
	    G^\mu_{\n \nu} = 8 \pi G \av{T^\mu_{\n \nu}}\ ,
	\end{equation}
	where
	\begin{equation}
        	\av{T^\mu_{\n \nu}}(t) = \frac{1}{T} \int^{t + T/2}_{t-T/2} T^\mu_{\n \nu}(t')\di t'\ .
	\end{equation}
	If the field is periodic, we can consider an integer number of periods as the integration interval. 
	However, similar results can be reached for fast-evolving  bounded solutions averaging over time spans
	much bigger than the inverse of its frequency but much smaller than the inverse of the expansion rate,
	$\omega^{-1} \ll T \ll \mathcal{H}^{-1}$. The averaging error in both cases results 
	$\mathcal{O}(\mathcal{H}T)$.
	
    To leading order we can drop the averages of total time derivatives, 	so it can be proved 
    \cite{Cembranos:2015oya} that
    \begin{equation}
        \av{\dot{\phi}^2/a^2} = -\av{\phi\ddot{\phi}/a^2} = \av{V'(\phi)\phi}
            + \mathcal{O}\Big(\frac{\mathcal{H}}{\omega_\text{eff}}\Big)\ , \label{eq:av_2}
    \end{equation}
    and with this result the effective equation of state can be written as
	\begin{equation}\label{eq:w_def}
	    w = \frac{\av{p}}{\av{\rho}} = \frac{\av{V'\phi-2V}}{\av{V'\phi + 2V}} = \frac{n - 2}{n + 2} 
            + \mathcal{O}\Big(\frac{\mathcal{H}}{\omega_\text{eff}}\Big)\ , 
	\end{equation}
	for power-law potentials $V(\phi)\propto \phi^n$. As it can be seen, a massive scalar field, 
	$V = m^2 \phi^2/2$, would behave as CDM. For this particular case we can solve the 
	equation of motion through a WKB expansion. Thanks to this adiabatic expansion in the parameter 
	$\mathcal{O}(\mathcal{H}/ma)$ we can perform the averages explicitly, isolating the fast evolving 
	factor and integrating by parts, as explained in 	\cite{Cembranos:2015oya}. For our model, we must 
	compute the first correction in $\lambda$. If the mass term is dominant, via a WKB expansion we 
	can write
	\begin{align}
	    \av{\phi^4} &\simeq \frac{3}{2}\av{\phi^2}\av{\phi^2}\ ,\\
	    \av{\rho} &\simeq m^2\av{\phi^2}\ ,
	\end{align}
	so the first anharmonic correction to the equation of state is
	\begin{equation}
	    w \simeq \frac{3\lambda}{8m^4}\av{\rho}\ .
	\end{equation}    
    In this effective description the background evolution of the field is described through its density
    and its effective equation of state, using the conservation equation
    \begin{equation}\label{eq:conservation}
        \dot{\rho}=-3\mathcal{H}(1+w)\rho\ .
    \end{equation}
    where for the equation of state $w$ we will use the formula
	\begin{equation}\label{eq:w_eff}
	    w = \frac{\displaystyle\frac{3\lambda}{8m^4}\rho}{\displaystyle 1+\frac{9\lambda}{8m^4}\rho}\ ,
	\end{equation}
	that smoothly interpolates between the radiation-like $w\simeq 1/3$ and matter-like $w\simeq 0$
	behaviour.
    Now we can apply the same trick to the evolution of the perturbations. The equations of motion for the
    fluid variables are
    \begin{align}
        \dot{\delta} &= 3\mathcal{H}(w-c_s^2)\delta - ku - \frac{1}{2}(1+w)\dot{h}\ ,\label{eq:pert_d_avg}\\
        \dot{u} &= -\mathcal{H}(1-3w)u + kc_s^2\delta\ ,\label{eq:pert_u_avg}
    \end{align}
    where $\delta\equiv \av{\delta\rho}/\av{\rho}$, $u\equiv (1+w)\av{v}$ stand for averaged quantities
    and $w$, $c_s^2$ are the effective equation of state and sound speed. To complete the system there
    only remains to compute the effective sound speed
    \begin{equation}
        c_s^2 = \frac{\av{\delta P}}{\av{\delta \rho}}\ .
    \end{equation}
    In contrast with the adiabatic sound speed, the sound speed is gauge-dependent. But, as we will 
    show now, the gauge ambiguities remain of order $\mathcal{O}(\mathcal{H}/\omega_\text{eff})$ so our
    effective sound speed turns out to be gauge-independent. In fact, identical expressions
    have previously been obtained working in the comoving gauge \cite{Li:2013nal} and in the Newtonian
    gauge \citep{Cembranos:2015oya}.
    To leading order we have
    \begin{align}
        \av{\partial_{\eta}(\dot{\phi}\delta\phi + \phi\delta\dot{\phi})} &= 0 
            + \mathcal{O}\Big(\frac{\mathcal{H}}{\omega_\text{eff}}\Big)\ . \label{eq:av_1}
    \end{align}
    Then, using \eqref{eq:sf_pert} and \eqref{eq:av_1} we can obtain the result
    \begin{equation}
        \av{\dot{\phi}\dot{\delta\phi}} = \frac{1}{2}\av{a^2V'\delta\phi + (k^2+a^2V'')\phi\delta\phi}
            + \Phi\av{a^2V'\phi} + \mathcal{O}\Big(\frac{\mathcal{H}}{\omega_\text{eff}}\Big)\ ,
    \end{equation}
    and finally compute the effective sound speed for a generic gauge
    \begin{align}
        c_s^2 &= \frac{\av{\delta P}}{\av{\delta \rho}}= 
            \frac{\frac{1}{2}\av{V'\delta\phi + ((k/a)^2+V'')\phi\delta\phi-2V'\delta\phi}
                -\Phi\av{\dot{\phi}^2/a^2-V'\phi}}
            {\frac{1}{2}\av{V'\delta\phi + ((k/a)^2+V'')\phi\delta\phi+2V'\delta\phi}
                -\Phi\av{\dot{\phi}^2/a^2-V'\phi}}
            + \mathcal{O}\Big(\frac{\mathcal{H}}{\omega_\text{eff}}\Big)\\
        &=\frac{\av{((k/a)^2+V'')\phi\delta\phi - V'\delta\phi}}
            {\av{((k/a)^2+V'')\phi\delta\phi + 3V'\delta\phi}}
            + \mathcal{O}\Big(\frac{\mathcal{H}}{\omega_\text{eff}}\Big)\ . \label{eq:cs2}    
    \end{align}
    As we anticipated, the gauge ambiguities in the metric perturbations remain of order 
    $\mathcal{O}(\mathcal{H}/\omega_\text{eff})$, so the final expression holds in any gauge.
    Moreover, it can be rewritten in a manifestly gauge-invariant form substituting $\delta\phi$ by its
    gauge-invariant perturbation \cite{Mukhanov:1990me}
    \begin{equation}
        \delta\phi^{\text{(gi)}} = \delta\phi + \dot{\phi}(B-\dot{E})\ ,
    \end{equation}
    and using the relations
    \begin{align}
        \av{V'\dot{\phi}} &= \av{\partial_\eta(V)} = 0 + \mathcal{O}\Big(\frac{\mathcal{H}}{\omega_\text{eff}}\Big)\ ,\\
        \av{V''\phi\dot{\phi}} &= \av{\phi\,\partial_\eta(V')} = -\av{V'\dot{\phi}} 
                +\mathcal{O}\Big(\frac{\mathcal{H}}{\omega_\text{eff}}\Big) 
            = 0 +\mathcal{O}\Big(\frac{\mathcal{H}}{\omega_\text{eff}}\Big)\ ,
    \end{align}
    we obtain
    \begin{equation}
        c_s^2 =\frac{\av{((k/a)^2+V'')\phi\delta\phi^{\text{(gi)}} - V'\delta\phi^{\text{(gi)}}}}
            {\av{((k/a)^2+V'')\phi\delta\phi^{\text{(gi)}} + 3V'\delta\phi^{\text{(gi)}}}}
            + \mathcal{O}\Big(\frac{\mathcal{H}}{\omega_\text{eff}}\Big)\ .
    \end{equation}
    This expression agrees with the result obtained in \cite{Cembranos:2015oya} working in the Newtonian
    gauge, so the same conclusions apply. In particular, a generic feature of this kind of models is
    a suppression of growth $c^2_s\simeq 1$ for small scales $k\gg \omega_\text{eff}$. In the case of 
    a power-law potential $V(\phi)=\frac{C}{n}\phi^n$, for large scales $k\ll \omega_\text{eff}$ we have
    \begin{equation}
        c^2_s = \frac{n-2}{n+2} + \mathcal{O}\Big(\frac{\mathcal{H}}{\omega_\text{eff}}\Big)\ .
    \end{equation}
    For a harmonic potential $n=2$, the zero-order term drops out and we must calculate the first-order
    corrections in $k$. Our potential of interest is a polynomial $V(\phi)=\frac{1}{2}m^2\phi^2 + 
    \frac{\lambda}{4}\phi^4$, a mass term plus an anharmonic correction, in this case we have
    \cite{Cembranos:2015oya}
    \begin{equation}
        c^2_s = \frac{k^2}{4m^2a^2} + \frac{3}{4}\frac{\lambda}{m^4}\rho\ ,
    \end{equation}
    where $\rho$ is the energy density of the scalar field and the anharmonic correction is assumed to 
    be small. In our numerical solution we will use an effective sound speed
    \begin{equation}\label{eq:cs2_eff}
	    c_s^2 = \frac{\displaystyle\Big(\frac{k}{2ma}\Big)^2+\frac{3}{4}\frac{\lambda}{m^4}\rho}
			{1+\displaystyle\Big(\frac{k}{2ma}\Big)^2+\frac{9}{4}\frac{\lambda}{m^4}\rho}\ ,
    \end{equation}
    suggested by the form of \eqref{eq:cs2} and that smoothly interpolates between all the regimes of 
    interest.

\section{Heuristic constraints on the non-harmonic contribution}\label{sec:heur}
    In this section we will discuss the simplest limits that constrain the model. With this objective let 
    us assume a simple cosmology composed of radiation, cosmological constant and our scalar field
    \begin{equation}
        \mathcal{H}^2 = a^2 H_0^2 \left( \Omega_\phi (a) 
            + \frac{\Omega_{\text{rad}}}{a^4} + \Omega_\Lambda\right)\ ,
    \end{equation}
    where $\Omega_i= 8\pi G \rho_i/(3 H_0^2)$ are the abundances with $i \in\{\phi, \text{rad}, 
    \Lambda\}$ which correspond to scalar field, radiation and cosmological constant respectively.    
    
\begin{figure}[t]
        \includegraphics[scale=0.6]{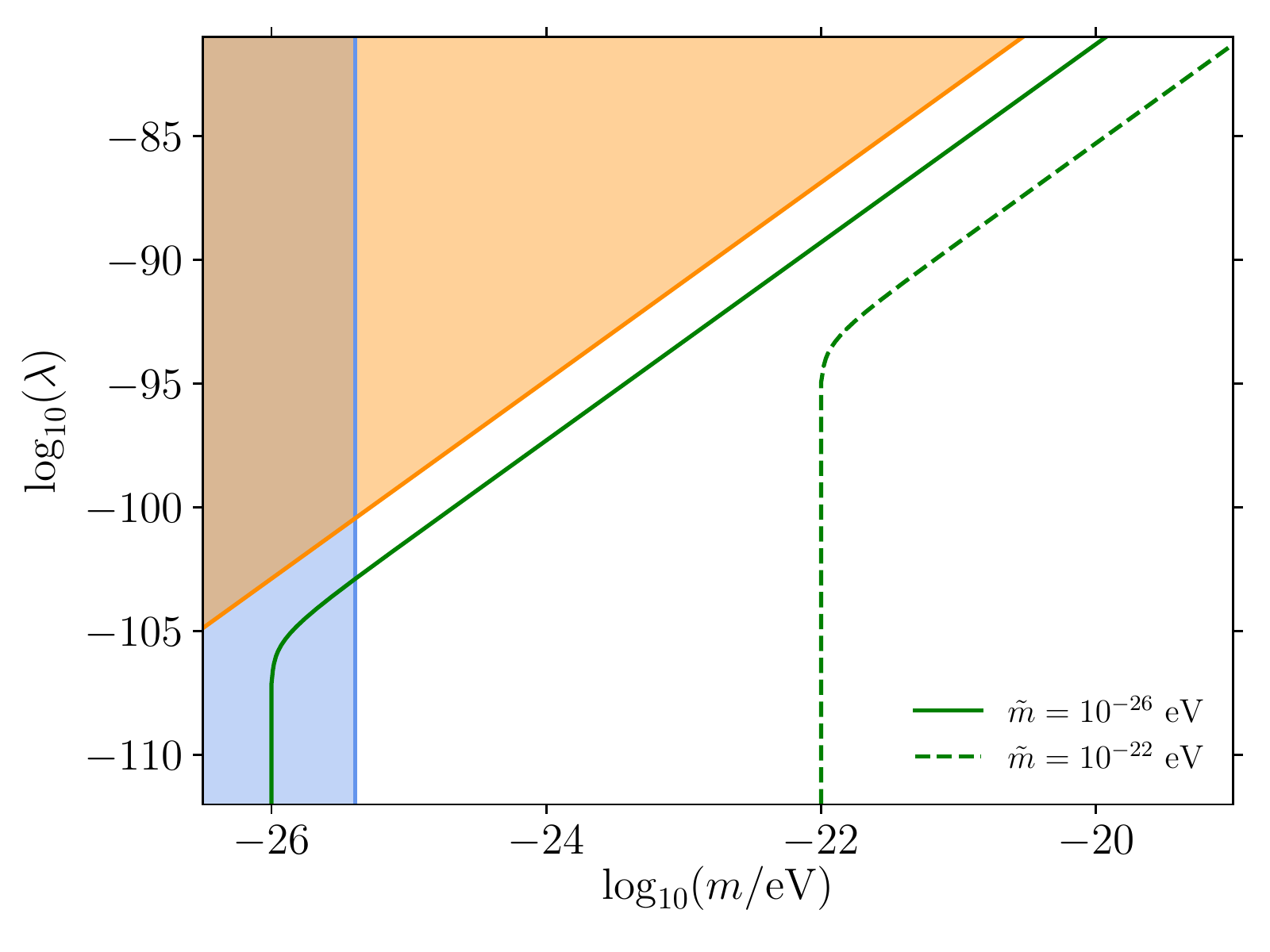}
        \caption{{\footnotesize Different heuristic bounds. Orange region
        corresponds to the parameter-space excluded for the effects of $\lambda$ on the background
        evolution. In the blue region, the effect of a non-negligible sound speed results in a strong
        disagreement with observations, hence it is excluded. The green curves represent 
        \eqref{eq:lambda_estimate} for two different masses. According to the argument in the main
        text, points along each curve should give similar structure-formation results.}}
        \label{fig:heuristic}
    \end{figure}

    \begin{itemize}
    \item
    \emph{Limits on $\lambda$ from background evolution.} 
        The position of the peaks in the CMB temperature spectrum, especially the first one, is very
        sensitive to the amount of matter and the redshift of equality $z_\text{eq}$. We can assume
        that to have a viable model of dark matter this quantities remain essentially the same as in
        $\Lambda$CDM. In this case, to have a dark matter behaviour that resemble CDM the
        anharmonic corrections at this time should be small
        \begin{equation}
            1\gg w \simeq \frac{3}{8}\frac{\lambda}{m^4}\rho_\phi(a_\text{eq})\ .     
        \end{equation}
        This imposes an upper limit on $\lambda$, namely
        \begin{equation}\label{eq:heuristic_limit_lambda}
            \lambda < \frac{8}{3}\frac{m^4}{\rho_\phi(a_\text{eq})}\ ,
        \end{equation}
        excluding the orange region in Figure \ref{fig:heuristic}.\\
    
    \item                    
    \emph{Limits on $m$ from perturbation evolution.} 
        If $\lambda$ is small enough, the background evolution of the effective fluid is identical to 
        $\Lambda$CDM. In this case we can obtain limits from the behaviour of the perturbations. From 
        \eqref{eq:pert_d_avg} and \eqref{eq:pert_u_avg} it can be seen that if we neglect the 
        expansion rate, $c_s^2k^2\gg\mathcal{H}^2$, density perturbations evolve according to 
        \begin{equation}
            \ddot{\delta} \simeq -c_s^2k^2\delta\ .
        \end{equation}
        producing an oscillatory behaviour instead of the standard growth.
        To avoid a clear disagreement with observations, the effect of a non-negligible sound speed must
        be small
        \begin{equation}\label{eq:sub_sound}
            c_{s}^2k^2 < \mathcal{H}^2\ .
        \end{equation}
        Translating into a lower bound in the allowed masses
        \begin{equation}
            m > \frac{k^2}{2a\mathcal{H}}\ ,
        \end{equation}
        as before, we assume that $z_\text{eq}$ corresponds to the standard value. To obtain a conservative
        limit we choose $k=0.2\ \text{Mpc}^{-1}$, the highest mode observed in LSS at the linear level,
        so that
        \begin{equation}
            m \gsim 10^{-26}\ \text{eV}\ ,
        \end{equation}
        excluding the blue region in Figure \ref{fig:heuristic}.\\        
    \item    
    \emph{Observable effects of anharmonic corrections.}
        Finally, there is a region in the parameter space that we cannot yet exclude and where the effects
        of anharmonic corrections to the sound speed may be important
        \begin{equation}
            c_s^2 \simeq \frac{k^2}{4m^2a^2}+ \frac{3}{4}\frac{\lambda}{m^4}\rho_\phi\ .
        \end{equation}
        Imposing that the second term dominates over the harmonic contribution yields an upper bound
        on $\lambda$
        \begin{equation}
            \lambda > \frac{k^2m^2}{3\rho_\phi a^2}\ ,
        \end{equation}       
        corresponding to a region that we cannot exclude right away, but where effects of the anharmonic 
        corrections to the sound speed are to be expected.\\
    \end{itemize}        
    
    An additional result that can be obtained from \eqref{eq:sub_sound} is the Jeans wavenumber
    \begin{equation}
        c_s^2k_J^2 = \mathcal{H}\ .
    \end{equation}
    Sub-Hubble modes below this Jeans wavenumber, $k < k_J$, grow while modes with $k>k_J$ are suppressed.
    In the massive case with $\lambda = 0$ we obtain
    \begin{equation}\label{eq:kjeans_mass}
        k_J^2 = 2a\mathcal{H}m\ .
    \end{equation}
    Now, since we have seen that the quartic correction affects the sound speed, it will also affect
    the Jeans scale. It is natural to ask what combination of parameters ($m$, $\lambda$) can have
    a similar impact on structure formation as the case ($\tilde{m}$, $\tilde{\lambda}=0$). To this end, we 
    look for the combination that gives the same Jeans scale at the matter-radiation equality. Since its
    scaling in time is not significantly modified, this simple estimate should capture the 
    essential features of structure formation in both models. Equating both sound speeds and inserting
    the result \eqref{eq:kjeans_mass} we have an estimate for $\lambda$
    \begin{equation}\label{eq:lambda_estimate}
        \lambda = 4.96\times 10^{-100}\,\Big(\frac{\tilde{m}}{10^{-24}\ \text{eV}}\Big)^3\Big(
            \frac{1-r^2}{r^4}\Big)\ ,\qquad r\equiv \frac{\tilde{m}}{m}\ .
    \end{equation}
    This simple result suggests for instance that, at the linear level, structure formation should be 
    similar in the models ($\tilde{m}=10^{-26}\ \text{eV}$, $\tilde{\lambda}=0$) and 
    ($m=10^{-24} \ \text{eV}$, $\lambda\simeq 4.96\times 10^{-98}$), a result that we will check with the 
    full numerical solution. This estimate is represented in Figure \ref{fig:heuristic} for two
    different masses $\tilde{m}$.
           
    After discussing some approximate bounds on our model and its physical origin, we will devote the
    next section to the full numerical solution.
    
\section{Numerical evolution and constraints}\label{sec:numeric}
\begin{figure}[htb]
		\subfigure{
		    \includegraphics[scale=0.5]{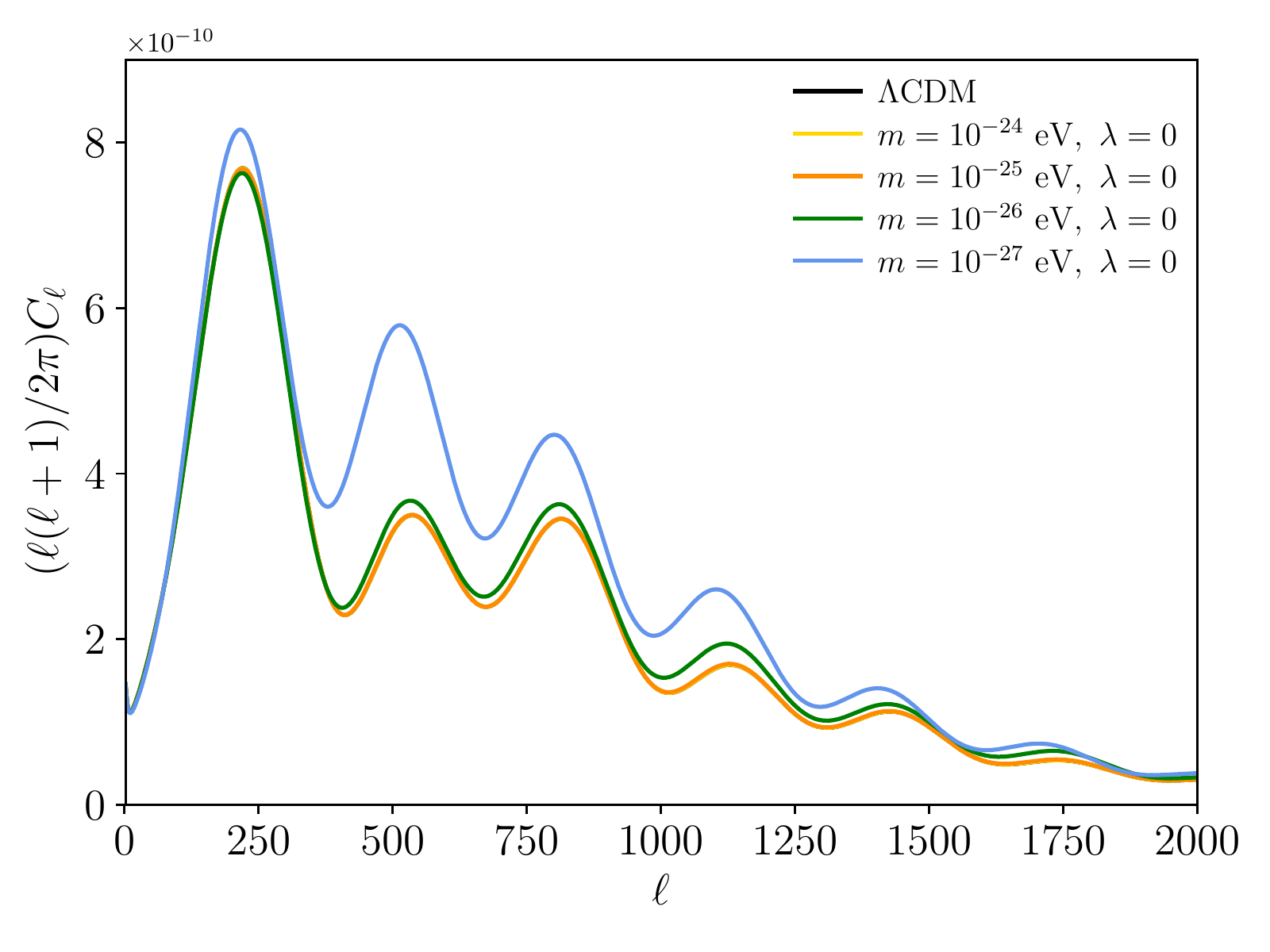}
		}
		\subfigure{
		    \includegraphics[scale=0.5]{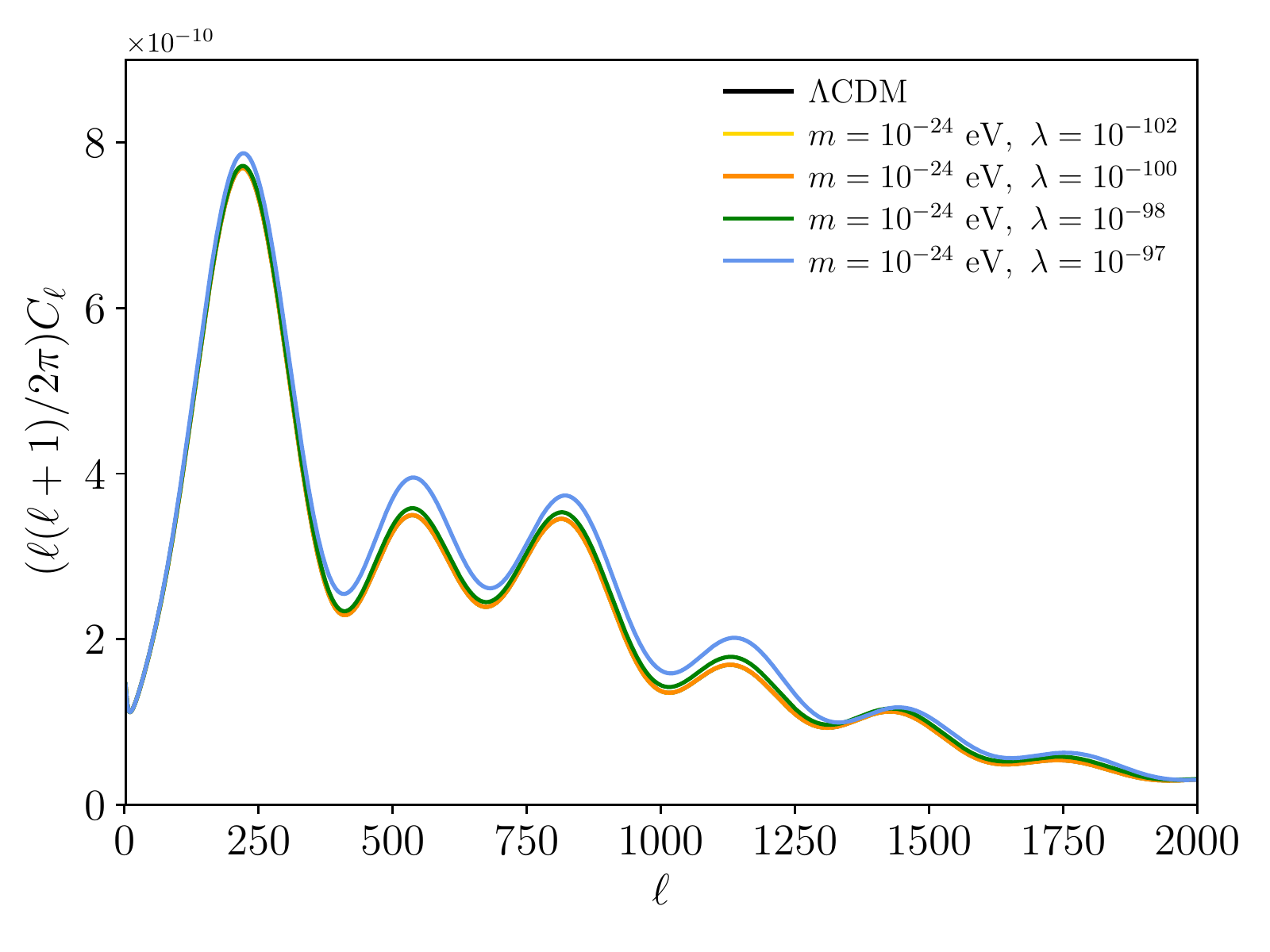}
		}
		\caption{Temperature power spectrum. On the left, results for a massive scalar field without
		         self-interaction. On the right, results for different self-interaction strengths for
		         a mass that is indistinguishable from CDM with $\lambda=0$.}
		\label{fig:cl}
    \end{figure}

\begin{figure}[htb]
		\subfigure{
		    \includegraphics[scale=0.5]{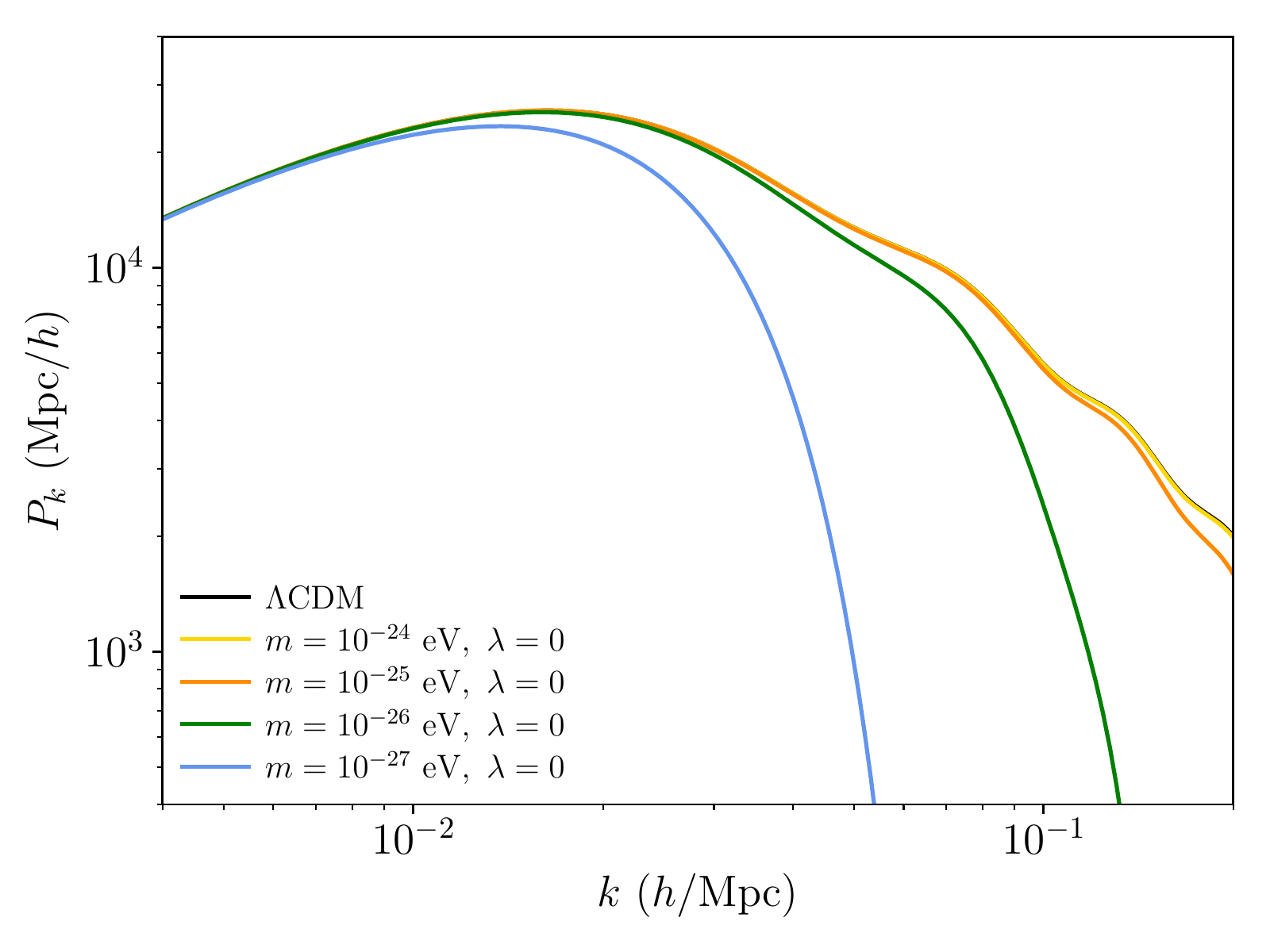}
		}
		\subfigure{
		    \includegraphics[scale=0.5]{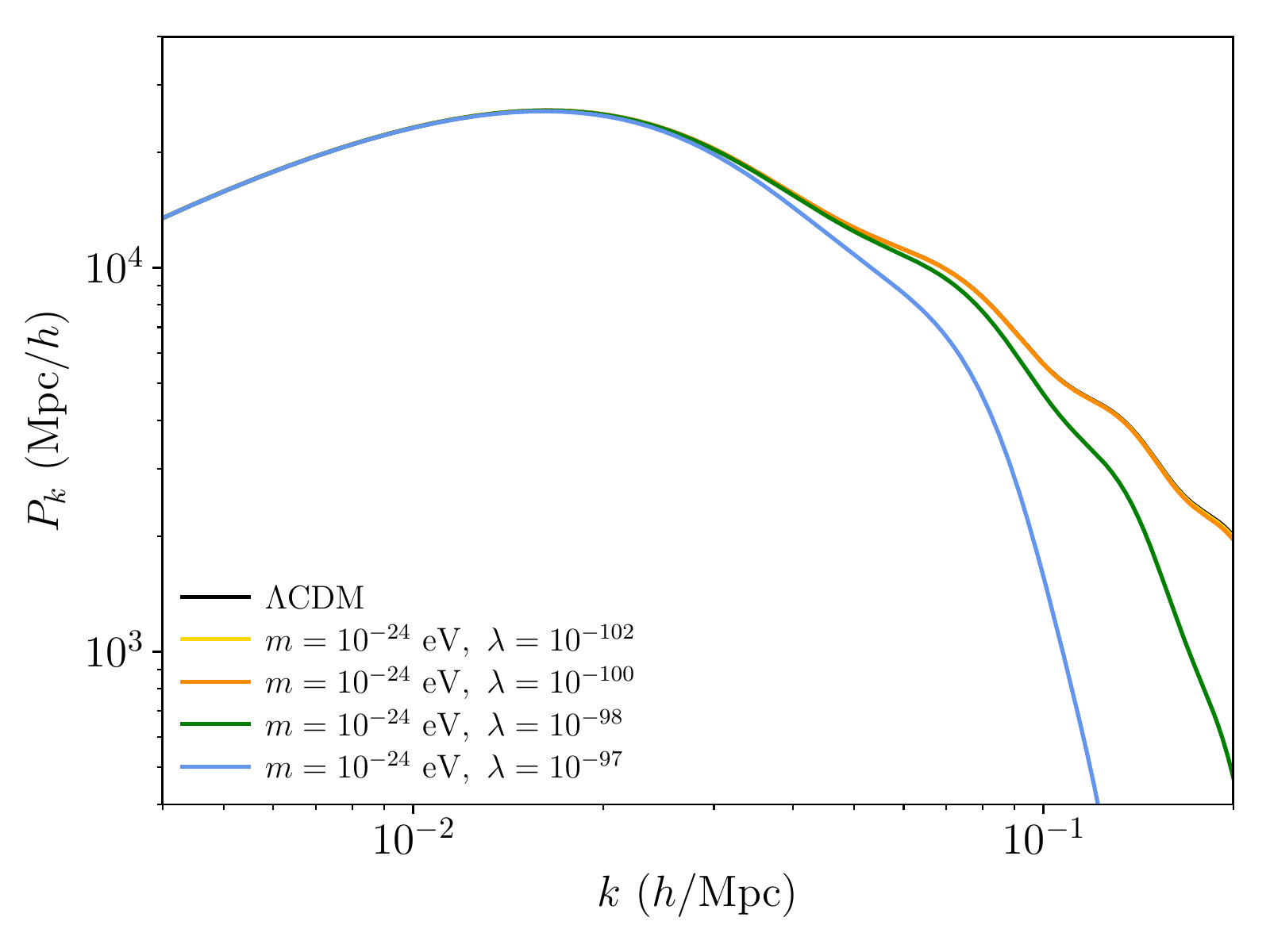}
		}
		\caption{Matter power spectrum. On the left, results for a massive scalar field without
		         self-interaction. On the right, results for different self-interaction strengths for
		         a mass that is indistinguishable from CDM with $\lambda=0$.}
		\label{fig:pk}
    \end{figure}

    We modify the publicly available Boltzmann code {\sc class} \citep{Blas:2011rf} and include this 
    ultralight scalar field as a new species, that will assume the role of dark matter. Now, we summarize 
    the key changes in the code and the evolution scheme chosen for the scalar field.
    \begin{itemize}
        \item At the background level, we start solving the equation \eqref{eq:sf_bg} with initial 
            conditions $\dot{\phi}=0$ and $\phi = \phi_0$. The initial value $\phi_0$ is chosen internally
            with a built-in shooting algorithm such as to match the energy density $\Omega_\phi$(today) 
            required. As a technical aside, it is critical to start with a sensible initial guess for
            $\phi_0$, so that the shooting algorithm converges quickly. In \citep{Hlozek:2014lca} the
            authors provide analytical formulae for the initial guess in the harmonic case, that works 
            as well if the anharmonic corrections are small. If the quadratic and quartic terms are 
            comparable it is more difficult to find analytical expressions that fit our purposes. In our
            case, we precompute an interpolation table for different values of $m$, $\lambda$ and $\phi_0$
            yielding some value $\Omega_\phi (m,\lambda,\phi_0)$. We only compute a coarse table, so that
            we still use the shooting algorithm to adjust $\phi_0$ and achieve the desired precision in
            $\Omega_\phi$.
            
            With the initial conditions provided, the field starts its evolution frozen, slowly rolling 
            down the potential until its natural frequency term in \eqref{eq:sf_bg} dominates and it 
            undergoes rapid oscillations. In this case it is computationally expensive to follow every
            oscillation so we turn to the averaged equations when $\sqrt{V''(\phi)}>3H$.
            
            In the averaged regime, we solve \eqref{eq:conservation}, matching continuously with the 
            solution in the exact regime, and compute the pressure using the effective equation of 
            state \eqref{eq:w_eff}.
          \item At the perturbation level, we first solve \eqref{eq:pert_d_exact} and \eqref{eq:pert_u_exact}
              with adiabatic initial conditions $\delta=u=0$. For each mode $k$ we start the integration
              early enough to ensure that we start well within the exact regime, 
              $\sqrt{V''(\phi)}\ll 3H$. In the averaged regime, $\sqrt{V''(\phi)}> 3H$, we solve the 
              equations \eqref{eq:pert_d_avg} and \eqref{eq:pert_u_avg} with the sound speed given by
              \eqref{eq:cs2_eff}.
    \end{itemize}
    Some results for temperature and matter power spectra are shown in Figures \ref{fig:cl} and
    \ref{fig:pk}. They show the impact of different choices of $m$ and $\lambda$, while the other 
    cosmological parameters are fixed to their Planck \citep{PlanckResults} best-fit values. As 
    anticipated, the main cosmological signature is the appearance of a cut-off in the matter power
    spectrum. This cut-off has already been discussed in the harmonic case \cite{Hlozek:2014lca}. In our
    case, we see that the anharmonic terms produce a similar effect.
    
    
    \subsection{Physical effects}
\begin{figure}[htb]
			\subfigure{
			    \includegraphics[scale=0.5]{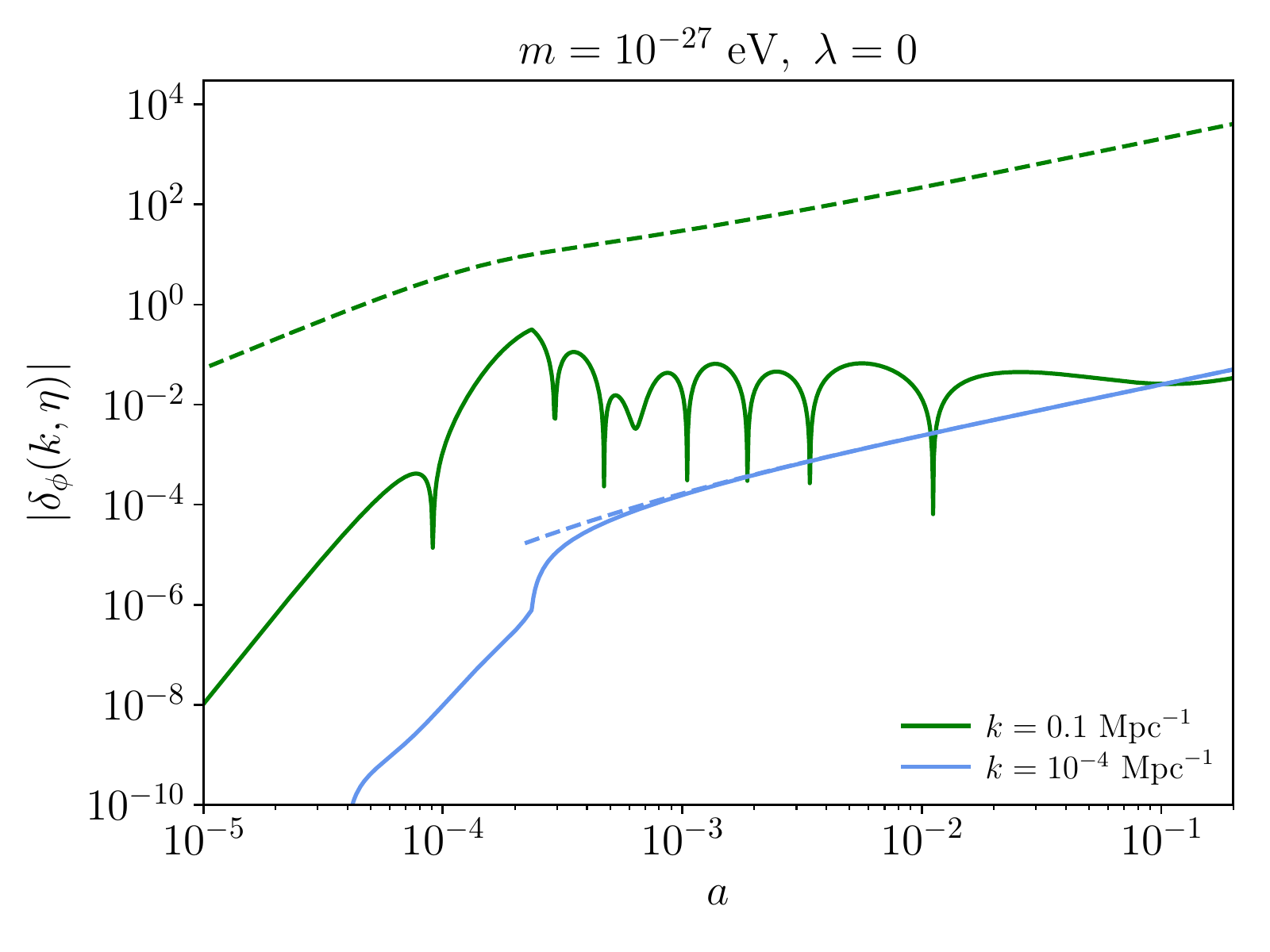}
			}
			\subfigure{
			    \includegraphics[scale=0.5]{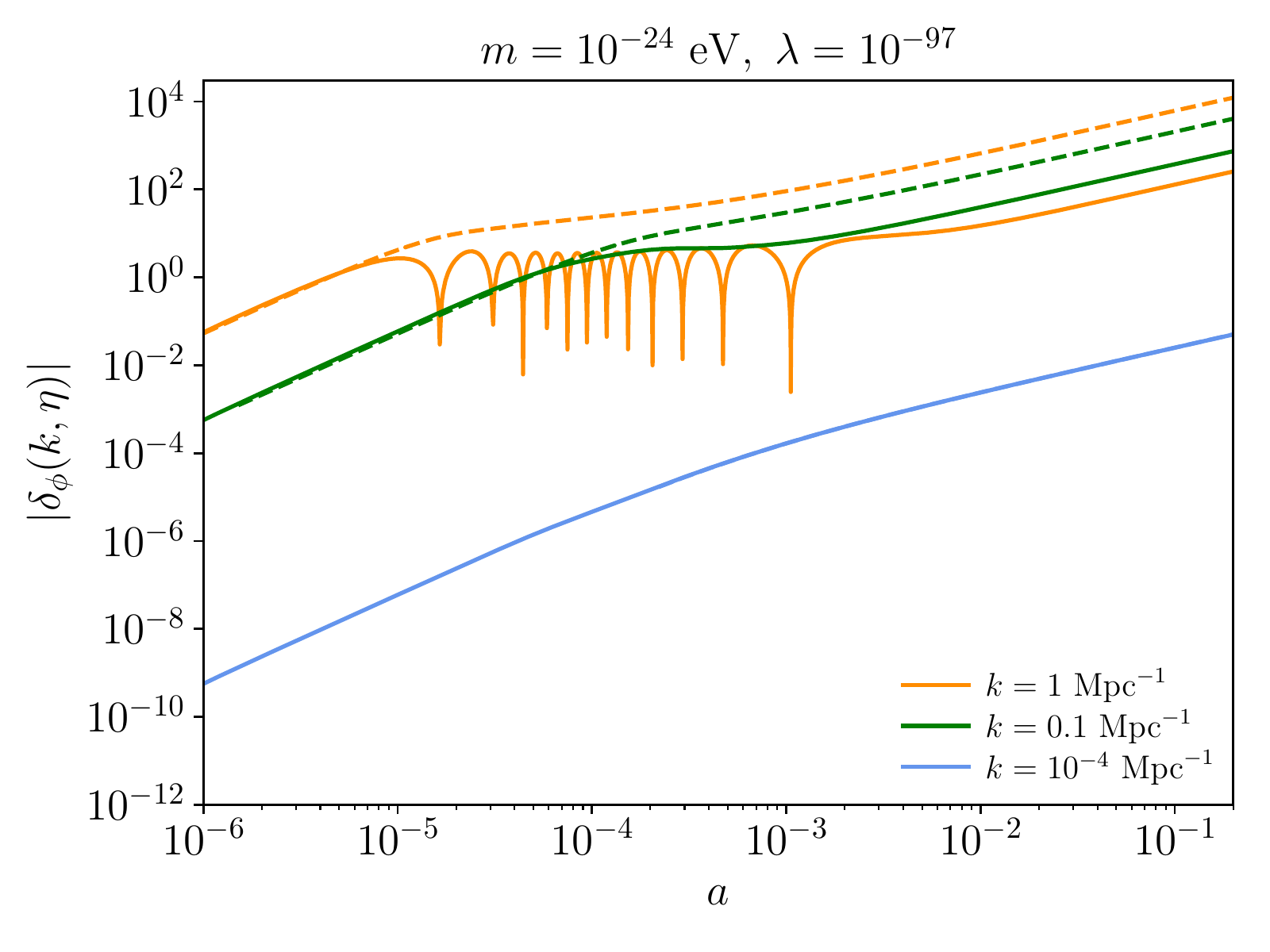}
			}
			\caption{Evolution in time of the dark matter transfer functions compared to the standard
			    $\Lambda$CDM evolution, represented by dotted lines.}
			\label{fig:matter_density}
	    \end{figure}

        The main physical effect responsible for the appearance of a cut-off in the matter power
        spectrum has already been discussed. In the averaged regime, the scalar field that supplies
        the dark matter component behaves like a fluid with a non-negligible sound speed. On small 
        scales, above a certain Jeans scale $k_J$, the density perturbations oscillate and the growth
        is suppressed. This effect is illustrated in the Figure \ref{fig:matter_density} for modes above
        and below $k_J$.\\
        
        In the case of the CMB temperature power spectrum, it is far more difficult to disentangle the
        physical effect responsible for each feature. We split the effects in two categories, those
        coming from the modified background evolution and those coming from the perturbations. 
        Furthermore, we will refer to two extreme cases $(m=10^{-27},\ \lambda=0)$ and 
        $(m=10^{-24},\ \lambda=10^{-97})$ as $m$-case and $\lambda$-case respectively. To gain some 
        insight into the CMB spectrum structure, we will rely on simplified, analytical estimates 
        \cite{Hu:1995en, Weinberg:2008zzc, Lesgourgues:2013qba} and work in the Newtonian gauge.
        
        \subsubsection*{Background evolution}
            The modified equation of state \eqref{eq:w_eff} changes the background evolution, modifying
            in particular the redshift of matter-radiation equality $z_\text{eq}$ and in general the 
            expansion history $a(\tau)$. In the $m$-case, the field transitions directly from the
            frozen value $w\simeq -1$ to a matter-like phase, while in the $\lambda$-case there is 
            an intermediate radiation-like phase. There are two key effects
            \begin{itemize}
                \item \emph{First peak position.} The position $\ell_\text{peak}$ of the first peak
                    can be estimated as 
                    \begin{equation}
				        \theta_\text{peak} = \frac{\pi}{\ell_\text{peak}}\simeq 
				            \frac{d_s\rvert_\text{dec}}{d_a\rvert_\text{dec}}\ ,
				    \end{equation}
				    where the angular diameter $d_a$ distance is defined as
				    \begin{equation}
				        d_a\rvert_\text{dec} = a_\text{dec}\int^{\eta_0}_{\eta_\text{dec}}\di\eta\ ,
				    \end{equation}
				    $d_s\rvert_\text{dec}$ is the sound horizon of the photon-baryon plasma 
				    evaluated at decoupling
				    \begin{equation}
				        d_s\rvert_\text{dec} = a_\text{dec} \int^{\eta_\text{dec}}_0c_{s\,\gamma}\di\eta\ ,
				    \end{equation}
				    and the sound speed for the baryon-photon plasma is
				    \begin{equation}\label{eq:cs_plasma}
				        c_{s\,\gamma}^2=\frac{1}{3(1+R)}\ ,\qquad R\equiv \frac{3\rho_b}{4\rho_\gamma}\ .
				    \end{equation}
				    The angular diameter distance is almost unaffected but the sound horizon is slightly
				    modified. Compared to $\Lambda$CDM we obtain relative deviations on $\ell_\text{peak}$
				    of about $+2\%$, shift to the left, in the $m$-case and $-0.7\%$, shift to the
				    right, in the $\lambda$-case. Both are compatible with the tiny deviations observed
				    in Figure \ref{fig:cl}.
				    
                \item \emph{Damping envelope.} Another physical scale that is modified is the diffusion
                    length
                    \begin{equation}
				        \lambda_D\rvert_\text{dec} = a_\text{dec}\Big(\int^{\eta_\text{dec}}_0 \Gamma^{-1} 
				            \di\eta\ \Big)^{1/2}\ ,
				    \end{equation}
				    where
				    \begin{equation}
				        \Gamma = an_e\sigma_Tx_e = -\dot{\kappa}\,x_e\ ,
				    \end{equation}
				    is the Thomson scattering rate. The diffusion length governs the damping envelope, 
				    $\text{e}^{-(\ell/\ell_D)^2}$, through the relation
				    \begin{equation}
				        \theta_D = \frac{\pi}{\ell_D}
				            \simeq \frac{\lambda_D\rvert_\text{dec}}{d_a\rvert_\text{dec}}\ .
				    \end{equation}
				    For a reference multipole $\ell=820$, corresponding to the third acoustic peak, 
				    in the $\lambda$-case we obtain a modified damping envelope that produces an 
				    enhancement of $6\%$ compared to $\Lambda$CDM, that can explain the overall increase
				    of power in Figure \ref{fig:cl}. For the $m$-case, we obtain the puzzling result
				    of a \emph{suppression} of $0.7\%$, in clear disagreement with the observed effect.
				    However, we will shortly see how a novel effect in the perturbation evolution can
				    account for this overall amplification.				   
            \end{itemize}
		\subsubsection*{Perturbation evolution}
		    In the tightly coupled regime, the photon temperature fluctuation evolves according to
		    \begin{equation}
		        \ddot{\Theta}_0 + \frac{\dot{R}}{1+R}\dot{\Theta}_0 + k^2c_{s\,\gamma}^2\Theta_0 =
		            -\frac{k^2}{3}\psi + \frac{\dot{R}}{1+R}\dot{\phi} + \ddot{\phi}\ ,
		    \end{equation}		
		    In the standard scenario, ignoring slow changes in $R$, $\phi$ and $\psi$ from the
		    expansion, we have
		    \begin{equation}
		        \ddot{\Theta}_0 + k^2c_{s\,\gamma}^2\Theta_0 \simeq -\frac{k^2}{3}\psi\ .
		    \end{equation}
		    This produces an oscillatory pattern with frequency $\omega = kc_{s\,\gamma}$ and zero-point displaced
		    by an amount $-(1+R)\psi$. The main part of the temperature Sachs-Wolfe effect comes 
		    from the contribution $|\Theta_0+\psi|^2\rvert_\text{dec}$, so the displacement of the 
		    zero-point of the oscillations gives the characteristic asymmetry between odd and even peaks
		    in the CMB temperature spectrum. Our modification of dark matter produces two interrelated
		    effects, oscillation and suppression of growth at small scales.
		    \begin{itemize}
		        \item \emph{Effects of suppression of growth at small scales.} The suppression of 
		            dark matter density perturbations at small scales also suppresses the gravitational 
		            wells $\psi$, shifting the zero-point of the oscillation back to zero.
		            This effect, alone, reduces the asymmetry among the peaks, decreasing
		            the odd and increasing the even peaks. This explains the characteristic enhancement
		            of the second peak with respect to the third one in Figure \ref{fig:cl}. 
		        \item \emph{Effects of oscillatory behaviour.} There only remains to explain one
		            effect, the striking gain in peak amplitude in the $m$-case. According to the
		            modification in the damping envelope, the peaks should be slightly suppressed 
		            and their enhancement is actually related to a resonance effect. In the standard
		            scenario, the term $\psi$ behaves like a constant external force, shifting the
		            equilibrium position of the photon oscillations. In our case, it is not constant
		            anymore, but oscillates with a frequency $kc_s$ given by the sound speed of the
		            dark matter perturbations \eqref{eq:cs2_eff}. These two frequencies, $kc_s$ and
		            $kc_{s\,\gamma}$, are comparable
		            for a range of $k$ values, as shown in Figure \ref{fig:sound_speed}, producing a 
		            resonant effect that increases the height of the peaks, as shown in Figure 
		            \ref{fig:photon_density}.
		            
                    \begin{figure}[htb]
						\includegraphics[scale=0.5]{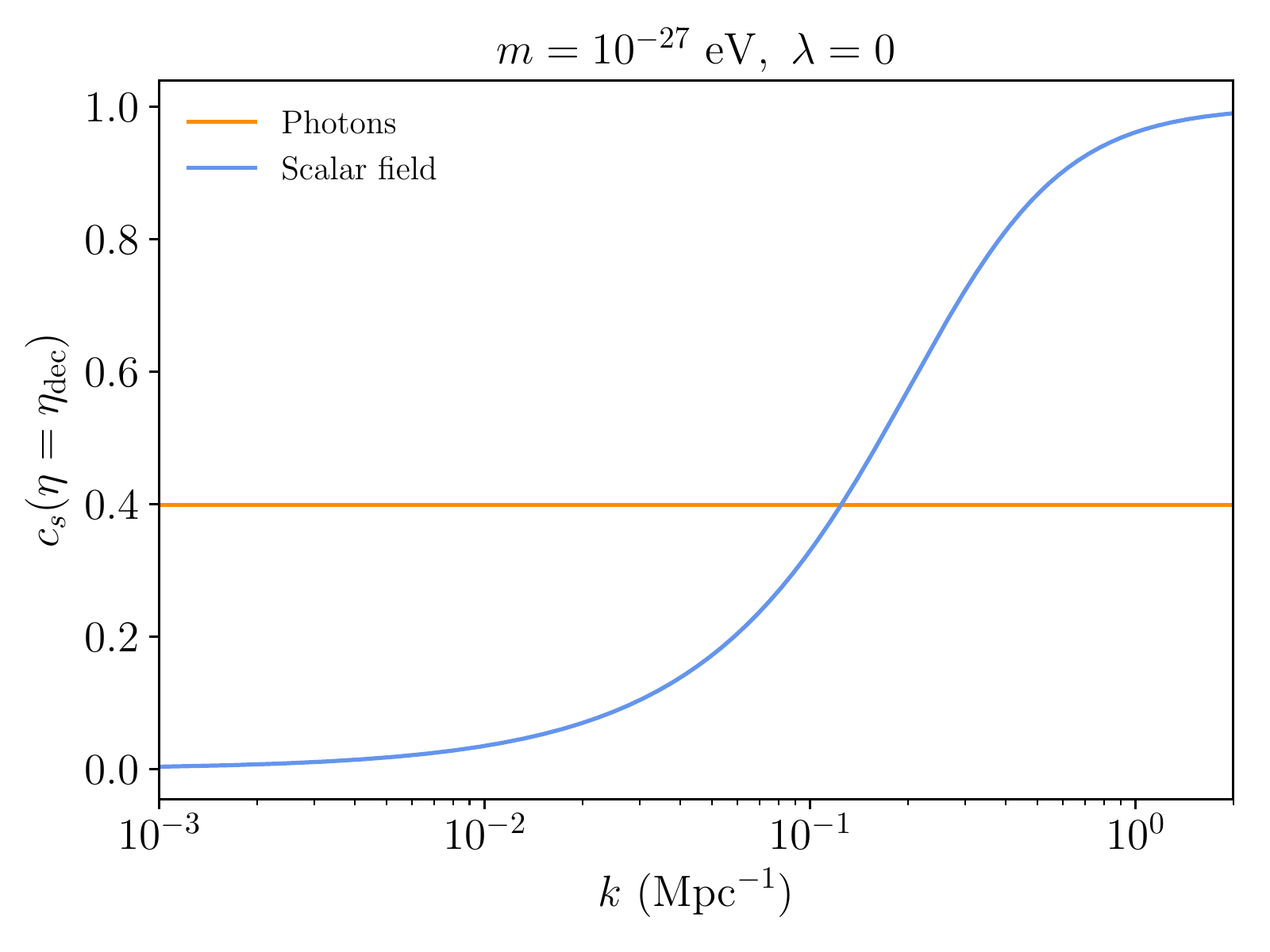}
						\caption{Sound speed at decoupling for photons and dark matter. Around
						    $k\simeq 0.1\ \text{Mpc}^{-1}$ the sound speed for both fluids, hence
						    the oscillation frequency too, are close and we have a resonant driving.}
						\label{fig:sound_speed}
				    \end{figure}
				    
				    \begin{figure}[htb]
						\includegraphics[scale=0.5]{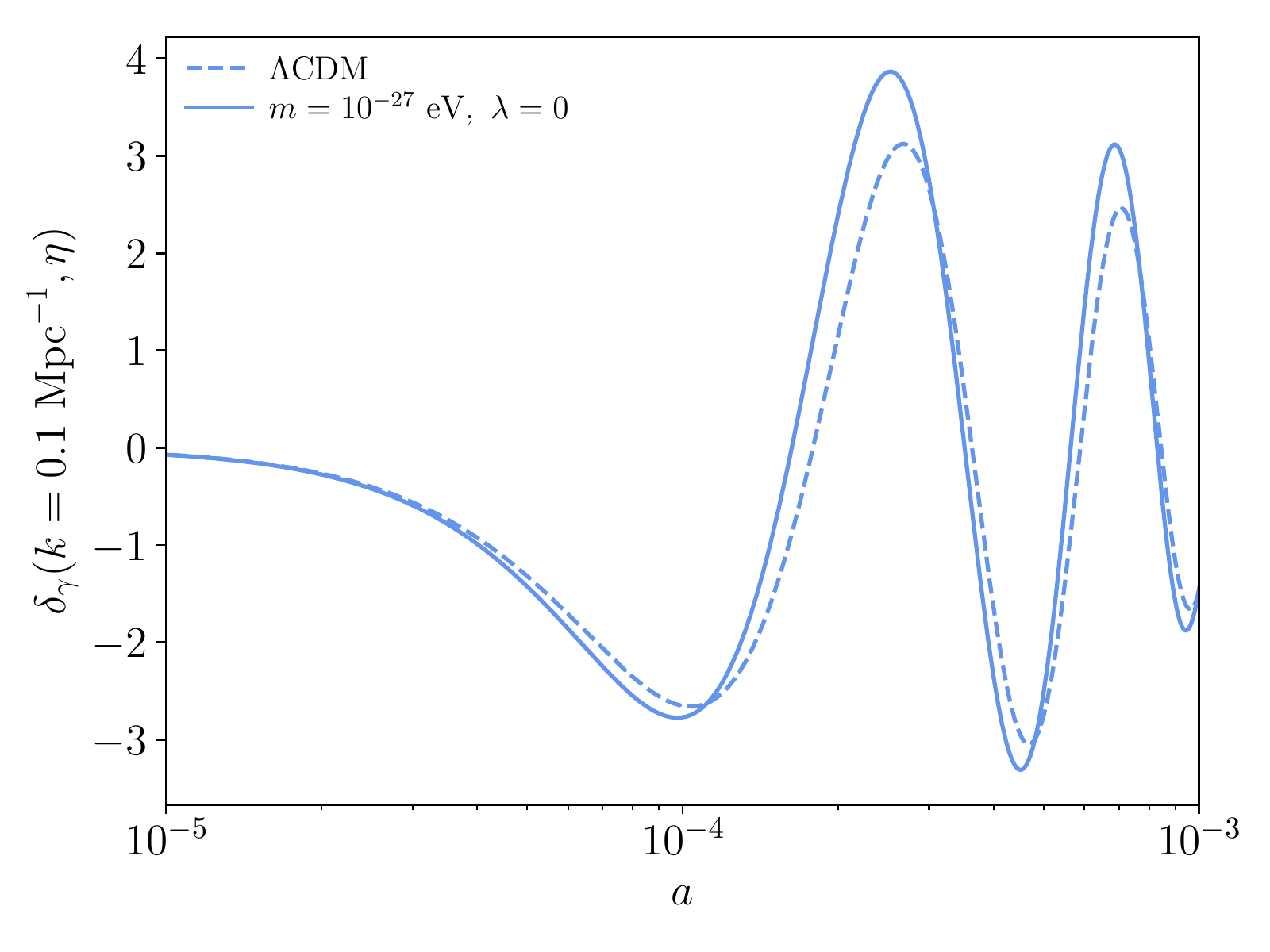}
						\caption{Evolution in time of the mode $k=0.1\ \text{Mpc}^{-1}$, corresponding
						    approximately to the fifth acoustic peak, until decoupling.}
						\label{fig:photon_density}
				    \end{figure}
		            
		            Moreover, since according to \eqref{eq:cs2_eff} the scale of the crossover in 
		            Figure \ref{fig:sound_speed} evolves $\propto a$, as we go from decoupling
		            back in time it moves to smaller $k$. 
		            That is to say, although the crossover at decoupling is located around 
		            $k\simeq 0.1\ \text{Mpc}^{-1}$, smaller $k$ have also fulfilled the resonance 
		            condition at previous times, so they have also got amplified.
		    \end{itemize}
    
    \subsection{Observational constraints}
        To compare this model with CMB and LSS observations and refine the heuristic constraints 
	    obtained in section \ref{sec:heur}, we use the public parameter-extraction code MontePython 
	    \citep{Audren:2012wb}. We will compare our results with two different data sets: CMB 
	    measurements by Planck and large-scale structure information by WiggleZ \cite{WiggleZResults}. We 
	    perform two analysis, Planck only and Planck+WiggleZ. In each case we vary the six $\Lambda$CDM base 
	    model parameters, in addition to the foreground parameters, plus $m$ and $\lambda$, the mass and 
	    anharmonic parameter. We choose logarithmic priors in our model parameters, as shown in
	    Table \ref{tab:priors}.
	    
	    \begin{table}
		\caption{\label{tab:priors} Prior ranges on the base $\Lambda$CDM parameters and the model
		    parameters $m$ and $\lambda$. A symbol $-$ means that there is no prior. Additionally, the fixed 
		    parameters include the neutrino properties. In our case, two massless neutrinos plus a 
		    massive one with $m=0.06\;\text{eV}$, such that $N_\text{eff}=3.046$ and 
		    $m_\nu/\Omega_\nu = 93.14\;\text{eV}$.}
		\begin{ruledtabular}
			\begin{tabular}{lcc}
			    Parameter & minimum & maximum\\
			\hline
			    $\Omega_bh^2$              & $-$      & $-$\\
			    $\Omega_{\phi}h^2$         & $-$      & $-$\\
			    $h$                        & $-$      & $-$\\
			    $\log(10^{10}A_s)$         & $-$      & $-$\\
			    $n_s$                      & $-$      & $-$\\
			    $\tau_\text{reio}^2$       & $0.04$   & $-$\\
			    $\log_{10}(m/\text{eV})$   & $-26$    & $-23.3$\\
			    $\log_{10}(\lambda)$       & $-111$   & $-98$\\
			\end{tabular}
		\end{ruledtabular}
		\end{table}
	    
	    It is important to note that to perform an accurate comparison with LSS data we
	    must restrict our analysis to linear scales $k\lsim 0.2\ h/\text{Mpc}$. The non-linear module in 
	    {\sc class} includes {\sc HaloFit} \cite{Bird:2011rb}, but since it has not been calibrated for
	    our model we restrict our analysis to linear scales without non-linear corrections. It is to be 
	    expected that, in the future, as more $N$-body simulations with ultralight fields become available, 
	    non-linear information will allow us to tighten the constraints.

\begin{figure}[htb]
            \includegraphics[scale=1]{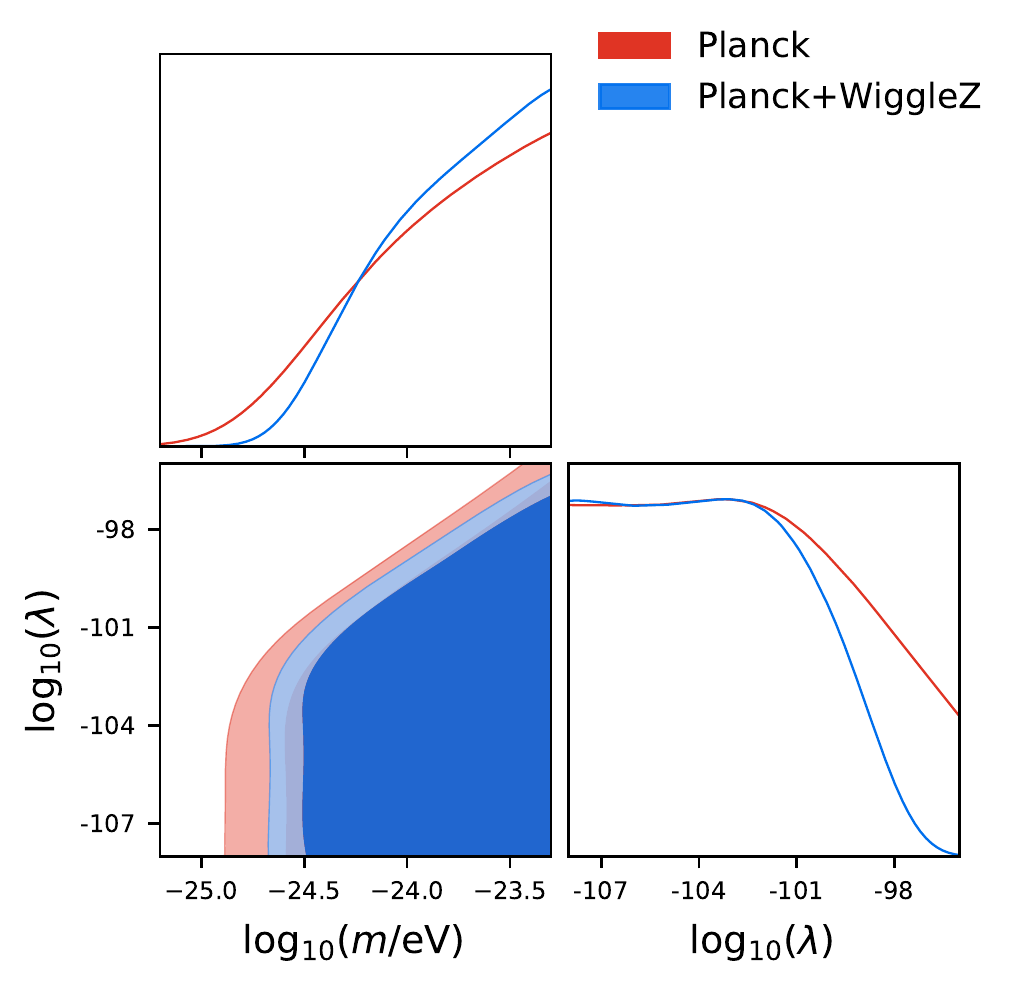}
            \caption{Contour plots with $95\%$ and $99\%$ confidence levels and 1d marginalized distributions. }
            \label{fig:contour}
        \end{figure}

        We do not observe any significant degeneracy between $m$, $\lambda$ and the rest of cosmological
        parameters. Best-fit results are shown in Table \ref{tab:bestfit}, while the marginalized countour 
        for our model parameters is represented in Figure \ref{fig:contour}.
        
        \begin{table}
		\caption{\label{tab:bestfit} Best fit results with 95$\%$ confidence level.}
		\begin{ruledtabular}
			\begin{tabular}{lcc}
			    Base parameters & Planck & Planck+WiggleZ\\
			\hline
			    $\Omega_bh^2$              & $0.02223\pm 0.00047$          & $0.02212^{+0.00042}_{-0.00041}$\\
			    $\Omega_{\phi}h^2$         & $0.1189^{+0.0044}_{-0.0041}$  & $0.1204^{+0.0032}_{-0.0034}$\\
			    $h$                        & $0.677\pm 0.019$              & $0.670^{+0.016}_{-0.014}$\\
			    $\log(10^{10}A_s)$         & $3.070^{+0.056}_{-0.053}$     & $3.057^{+0.046}_{-0.041}$\\
			    $n_s$                      & $0.965^{+0.016}_{-0.021}$     & $0.963^{+0.011}_{-0.010}$\\
			    $\tau_\text{reio}^2$       & $0.070^{+0.028}_{-0.029}$     & $0.061^{+0.024}_{-0.021}$\\
			    $\log_{10}(m/\text{eV})$   & $> -24.5$                     & $> -24.4$\\
			    $\log_{10}(\lambda)$       & $-$                           & $< -99.0$\\
			\hline
			\hline
			    Derived parameters     &&\\
			\hline
			    $z_\text{reio}$            & $9.2^{+2.6}_{-2.7}$             & $8.4^{+2.2}_{-2.1}$\\
			    $\Omega_\Lambda$           & $0.690^{+0.024}_{-0.027}$       & $0.681\pm 0.021$\\
			    $Y_\text{He}$              & $0.24778\pm 0.00020$            & $0.24773\pm 0.00018$\\
			    $100\theta_s$              & $1.04193^{+0.00098}_{-0.00099}$ & $1.04182^{+0.00084}_{-0.00083}$\\
			\end{tabular}
		\end{ruledtabular}
		\end{table}        
        
\section{Conclusions}\label{sec:conclusions}
    The presence of self-interactions in the ultralight field potential can lead to the appearance of
    new background-evolution phases, like the radiation-like due to our quartic potential. This modified 
    background evolution, and especially its critical effect on the sound speed of dark matter 
    perturbations, can lead to significant differences from observations. The observational signatures of 
    the anharmonic contribution are similar to the mass term, the most prominent being the appearance of a 
    cut-off in the matter power spectrum. This produces constraints for masses that would be otherwise 
    indistinguishable from CDM, i.e. $m\gsim 10^{-24}\;\text{eV}$. Our constraints on $\lambda$ complement 
    other bounds present in the literature, e.g. \cite{Dev:2016hxv}. This bounds on $\lambda$ follow a 
    scaling law with $m^4$ according to \eqref{eq:heuristic_limit_lambda}. We can extrapolate the results
    to higher masses using the $2\sigma$ region of Figure \ref{fig:contour}, obtaining an approximate
    constraint on $\lambda$
    \begin{equation}\label{eq:fitting_limit}
        \log_{10}(\lambda) < -91.86 + 4\log_{10}\Big(\frac{m}{10^{-22}\ \text{eV}}\Big)\ ,
    \end{equation}
    for masses $m > 10^{-24}\ \text{eV}$.
    
    So far, we have only analyzed linear observables, but in fact larger effects on non-linear 
    scales are expected. The available parameter space could be further constrained in the future using 
    cosmological information with non-linear observables, as more simulations with ultralight fields 
    become available. Even without non-linear information, using the formula \eqref{eq:lambda_estimate} 
    one could put forward the proposal that similar results on structure formation could be obtained for 
    higher masses with a positive $\lambda$. For instance, results for $\tilde{m}=10^{-22}\ \text{eV}$ 
    might be reproduced with masses $m\simeq 10^{-5}\ \text{eV}$ adding a self-interaction of the order 
    of $\lambda\simeq 10^{-24}$, very close to the limit that can be obtained from \eqref{eq:fitting_limit}. 
    Nevertheless, a definitive answer to this suggestive proposal require a fully non-linear analysis.
    
\vspace{0.2cm}    
{\bf Acknowledgements:}
This work has been supported by the MINECO (Spain) projects FIS2014-52837-P, FIS2016-78859-P(AEI/FEDER, UE),
 and Consolider-Ingenio MULTIDARK CSD2009-00064.

\bibliography{Biblio_scalar}

\end{document}